# Understanding ELM mitigation by resonant magnetic perturbations on MAST

A.Kirk, I.T. Chapman, Yueqiang Liu, P. Cahyna[2], P. Denner[3], G. Fishpool, C.J. Ham, J.R. Harrison, Yunfeng Liang[3], E. Nardon[4], S. Saarelma, R. Scannell, A.J. Thornton and the MAST team

EURATOM/CCFE Fusion Association, Culham Science Centre, Abingdon, Oxon, OX14 3DB, UK
[2]Institute of Plasma Physics AS CR v.v.i., Association EURATOM/IPP.CR, Prague, Czech Republic
[3]EURATOM-FZ Julich, D-52425 Julich, Germany
[4]Association Euratom/CEA, CEA Cadarache, F-13108, St. Paul-lez-Durance, France

## Abstract

Sustained ELM mitigation has been achieved using RMPs with a toroidal mode number of n=4 and n=6 in lower single null and with n=3 in connected double null plasmas on MAST. The ELM frequency increases by up to a factor of eight with a similar reduction in ELM energy loss. A threshold current for ELM mitigation is observed above which the ELM frequency increases approximately linearly with current in the coils. A comparison of the filament structures observed during the ELMs in the natural and mitigated stages shows that the mitigated ELMs have the characteristics of type I ELMs even though their frequency is higher, their energy loss is reduced and the pedestal pressure gradient is decreased. During the ELM mitigated stage clear lobe structures are observed in visible-light imaging of the X-point region. The size of these lobes is correlated with the increase in ELM frequency observed. The RMPs produce a clear 3D distortion to the plasma and it is likely that these distortions explain why ELMs are destabilised and hence why ELM mitigation occurs.

## 1. Introduction

All current estimations of the energy released by type I ELMs indicate that, in order to ensure an adequate lifetime of the divertor targets on ITER, a mechanism is required to decrease the amount of energy released by an ELM, or to eliminate ELMs altogether [1]. One such amelioration mechanism relies on perturbing the magnetic field in the edge plasma region, either leading to more frequent smaller ELMs (ELM mitigation) or ELM suppression. This technique of Resonant Magnetic Perturbations (RMPs) has been employed on DIII-D [2][3] and KSTAR [4], where complete ELM suppression has been possible, and on JET [5], MAST [6] and ASDEX Upgrade [7] where ELM mitigation has been obtained. In the mitigated phase on ASDEX Upgrade the type I ELMs are replaced by small-scale and high-frequency edge perturbations, which resemble small ELMs [8]. So although it is not the complete ELM suppression, it could be referred to a type I ELM suppression.

Although ELMs pose a threat to the divertor lifetime they do reduce the impurity content of the core plasma [9][10]. Recent calculations have shown that to avoid tungsten accumulation in the ITER core plasma an ELM frequency of greater than 30 Hz is required [11]. Although the ELMs associated with such an ELM frequency are predicted to be below the damage threshold for the divertor target [11], they would still produce a large thermal cycling of the target materials. ELM suppression would be the best solution for the divertor as long as the impurity transport associated with it is sufficient to avoid tungsten accumulation.

In this paper the effects of applying RMPs to MAST type I ELMing H-mode plasmas will be presented. MAST is equipped with 18 RMP coils (6 in the upper row and 12 in the lower row). These coils give considerable flexibility since they not only allow, for the first time, higher toroidal mode numbers (n=4 and n=6) but also improved alignment of the magnetic perturbations with the plasma equilibrium by allowing the phase of the applied field to be varied during the shot. In section 2 examples of ELM mitigation will be presented. Section 3 reports on an attempt to use vacuum and plasma response modelling to try to find a parameter that can determine the onset of ELM mitigation. Section 4 shows the effect the RMPs have on the pedestal and the implications for edge stability. Section 5 reports on the distortions to the plasma boundary near to the X-point produced by RMPs and finally section 6 summarises the observations and discusses their implications for ELM mitigation.

## 2. Examples of ELM mitigation in LSND and CDN

ELM suppression has not been achieved on MAST but ELM mitigation has been established in a range of plasmas using RMPs with toroidal mode numbers of n = 3, 4 or 6. ELM mitigation has been established in a Lower Single Null Divertor (LSND) and Connected Double Null (CDN) magnetic configurations using either a single or double row of coils. Figure 1a shows the poloidal cross section for typical LSND and CDN discharges. Due to the up-down symmetry in the divertor coils on MAST, LSND discharges are usually produced by shifting the plasma downwards. In this configuration the plasma is far from the upper row of RMP coils and hence the perturbation is predominantly from the lower row of coils, which produces a much broader spectrum of magnetic perturbation.  The CDN discharge is equally spaced from both rows of coils.

Figure 1b-e shows an example of the effect of the application of the RMPs with a toroidal mode number n=6 from the lower row of coils on a LSND H-mode plasma, which has a plasma current ($I_P$) of 600 kA, a toroidal magnetic field ($B_T$) of 0.55 T at a radius of 0.8 m (giving an edge safety factor ($q_{95}$) of 2.8), a line average density ($\bar{n_e}$) of $4 \times 10^{19}$ m$^{-3}$, and heated by 3.6 MW of Neutral Beam Injected ($P_{NBI}$) power.  Figure 1e shows the $D_\alpha$ trace for the case where the RMPs are applied with 5.6 kAt in an n=6 configuration.  The application of the RMPs produces an increase in ELM frequency ($f_{ELM}$) to ~ 200 Hz, which is a factor 3 increase over the natural ELM frequency and a consequent decrease in ELM size and line average density.

Figure 1f-g shows an example of the effect of the application of the RMPs with a toroidal mode number n=3 from both rows of coils in an even parity configuration on a CDN H-mode plasma, which has $I_P$ = 600 kA,  $B_T$ = 0.585T, $q_{95}$ = 7.2,  $\bar{n_e}$ = $3 \times 10^{19}$ m$^{-3}$ and $P_{NBI}$= 3.2 MW.  Figure 1i shows the $D_\alpha$ trace for the case where the RMPs are applied with 5.6 kAt in an n=3 configuration.  The application of the RMPs produces one of the largest increases in $f_{ELM}$ (~ 900 Hz), which represents a factor of 8 increase over the natural ELM frequency.

In order to maximise the increase in ELM frequency and to get a better understanding of the parameters determining why the RMPs are having an effect, dedicated scans have been performed.  Discharges have been repeated with increasing current in the coils ($I_{ELM}$) to determine the threshold current for the onset of ELM mitigation together with the effect on ELM frequency. In the case of the LSND discharges, n=4 and n=6 RMP configurations have been used, while for the CDN discharge an n=3 configuration with the coils either in the even (where the currents in the upper and lower coil at the same toroidal location have the same sign) or 90L (where the current in the upper coil has the same sign as a lower coil at a toroidal location displaced by 90 degrees) have been used. Figure 2a shows the increase in the normalised ELM frequency (defined as the mitigated ELM frequency divided by the natural ELM frequency) as a function of $I_{ELM}$. In all cases the ELM frequency is observed to increase approximately linearly with $I_{ELM}$ above a threshold value, which is configuration specific.

Another way of varying the size of the applied RMP is to vary the distance of the plasma from the coils. Figure 2b shows the resulting normalised ELM frequency from a series of LSND discharges with the coils in an n=6 configuration and CDN discharges with the coils in an n=3 even parity configuration with different distances between the plasma and the coils. The ELM frequency clearly changes with distance to the coils with a clear threshold separation in the two cases. The maximum increase in ELM frequency obtained is a factor of 9 shown for CDN discharge with RMPs in the n=3 even configuration.

The sensitivity of the ELM frequency to the alignment of the applied perturbation with the pitch of the equilibrium magnetic field (i.e. to test if a resonant condition exists) can be performed on MAST either by changing the equilibrium (a $q_{95}$ scan) or by changing the angle of the applied field (a pitch angle scan). While the $q_{95}$ scan is traditionally used on devices to look for a resonant condition (see for example [3]) it has the disadvantage that it changes the underlying plasma parameters. In an n=3 configuration MAST can exploit a unique capability allowed by having 12 coils in the lower row.  If the current is kept fixed in the 6 upper coils, then by operating the lower coils in pairs and by varying the relative current in each pair the pitch angle of the applied field can be changed.  This allows a resonance condition to be studied without changing the underlying plasma parameters.

Figure 3a  shows the results from repeat LSND discharges with $B_T$ in the range 0.48 to 0.585 T, corresponding to $q_{95}$ in the range 2.4 to 3.0.  For this range of $B_T$ there is very little change in natural ELM frequency, however, both the n=4 (performed with $I_{ELM}$ = 4.0 kAt) and n=6 (performed with $I_{ELM}$=5.6 kAt) perturbations from the lower row of coils show a dependence of $f_{ELM}$ on $q_{95}$.  Figure 3b shows the resulting ELM frequency from repeat CDN discharges where the pitch angle ($\alpha$) of the applied RMPs has been varied. $\alpha$ is defined as the angle between the centre of the upper coil and the centroid of the paired lower coils with the same sign, chosen such that the angle is nearest to the pitch angle of the plasma equilibrium field lines at $\Psi_N$=0.95 at the LFS. Repeat discharges were performed with different pitch angles and the ELM frequency determined for each.  Again $f_{ELM}$ is found to be dependent on the alignment of the applied field.

While clear ELM mitigation has been observed, ELM suppression has not been established.  Since ELM suppression has been established in DIII-D at high and low collisionality [2][3], while only ELM mitigation is observed at intermediate values, and noting that there is a density threshold for complete suppression of type I ELMs on ASDEX Upgrade [7], a scan in fuelling rate and density has been performed.  The lower limit for the scan range is set by the minimum density required to achieve H-mode at the available heating power while the upper limit is set by the maximum density that can be achieved whilst maintaining the plasma in a type I ELM-ing regime. For the LSND discharges, the collisionality range scanned (Figure 4a) coincides with the window for which DIII-D do not observe ELM suppression ($0.3 < \nu_e^* < 2.0$). On ASDEX Upgrade the suppression of type I ELMs is not associated with collisionality, but rather the plasma density expressed as a fraction of the Greenwald density ($n_{GW}$), with suppression being observed for ne/$n_{GW}$>0.53 [8]. Figure 4b shows the distribution of the Greenwald density fraction for the MAST LSND discharges. While most lie in the range $0.2 < n_e/n_{GW} < 0.4$, a few discharges have been performed in the range that overlap with the ASDEX Upgrade type I ELM suppression region. For the CDN discharges both the collisionality and the density as a fraction of Greenwald number fall into the region that type I ELM suppression has been observed on DIII-D and ASDEX Upgrade. However, it should be noted that while ELM suppression has been established in LSDN on both devices, the access conditions for ELM suppression in CDN discharges in uncertain. Since in the only results published to date, DIII-D have reported that it was not possible to obtain ELM suppression in CDN discharges [12].

### 3. Vacuum and plasma response modelling

The ERGOS code (vacuum magnetic modelling) [13] has been used to calculate the magnetic perturbations to the plasma due to the coils. Its implementation on MAST has been previously described in reference [14]. Figure 5a shows a plot of the normalised ELM frequency as a function of the maximum value of the radial field component normalised to the toroidal magnetic field ($b^r_{res}$) calculated from ERGOS for the discharges used in the $I_{ELM}$ scan shown in Figure 2a.

Only the dominant toroidal harmonic is considered (i.e. n=3, 4 or 6) and other harmonics are neglected. In all cases the normalised ELM frequency increases linearly with $b^r_{res}$ above a threshold value. Although the threshold is similar for the n=4 SND, and n=3 even and 90L configurations ($b^r_{res}$(thresh) = $0.4x10^{-3}$) it is different for the n=6 SND configuration ($b^r_{res}$(thresh) = $0.7x10^{-3}$).

The discharges used in the $\Delta R_{coil}$ scan shown in Figure 2b have also been modelled using ERGOS. Figure 5b shows that the normalised ELM frequency increases linearly above a threshold value of $b^r_{res}$ of $0.5x10^{-3}$ and $1.1x10^{-3}$ for the n=3 (CDN even) and n=6 (LSND) cases respectively. This threshold value is different to that found during the $I_{ELM}$ scan and in the n=3 CDN case the rate of increase of ELM frequency with $b^r_{res}$ is more rapid.

Calculations have been performed using the MARS-F code, which is a linear single fluid resistive MHD code that combines the plasma response with the vacuum perturbations, including screening effects due to toroidal rotation [15]. The resistive plasma response significantly reduces the field amplitude near rational surfaces and reduces the resonant component of the field by more than an order of magnitude. The MARS-F calculations have been performed for the discharges used in the $I_{ELM}$ scan shown in Figure 2a and the value of $b^r_{res}$ taking into account the plasma response has been calculated. Figure 6 shows the normalised ELM frequency versus $b^r_{res}$ with the plasma response included. Similar to what was observed in the vacuum calculations, there is a different threshold in all of the cases and hence it has not been possible to identify a single parameter that can determine the onset of ELM mitigation.

The toroidal rotation velocity ($V_\phi$) has been measured using charge exchange recombination spectroscopy in a series of LSND discharges to which the RMPs have been applied in n=3, 4 and 6 configurations. In each case the RMPs produce a braking of the toroidal rotation, which in the case of n=3 is so severe that it produces a back transition to L-mode before any sustained ELM mitigation can be achieved. The lines in Figure 7a show the core value of $V_\phi$ (measured at normalised poloidal flux $\Psi_{pol} = 0.3$) as a function of time after the RMPs have been applied. In each case the deceleration of the plasma is similar, with just the saturated level being different for the three cases i.e. $V_\phi^{min} = 0$ (n=3), 10 (n=4) and 30 (n=6) $kms^{-1}$. The quasi-linear MARS-Q code [16] has been used to simulate the RMP penetration dynamics and the toroidal rotation braking for these shots. The model includes both the $\vec{j} \times \vec{B}$ [17] and the NTV torque [18] with both the resonant and non-resonant contributions included. The results of the simulations are shown as the symbols in Figure 7a. For the n=3 configuration the code predicts a full damping of the toroidal rotation, which is initially due to the $\vec{j} \times \vec{B}$ torque, in a time of less than 40ms, very similar to what is observed in the experiment. For the n=4 and n=6 configurations a similar rate of damping is predicted together with a prediction that a minimum saturated level would be achieved. The value of the saturated level is in good agreement for the n=6 configuration but the code predicts a higher saturated level for the n=4 configuration than what is observed experimentally. This could be due to the fact that the simulation only includes the n=4 component of the applied field whereas the n=4 coil configuration also has a sizeable n=8 sideband.

In contrast to the LSND discharges, for the CDN discharges presented here, little braking of the core plasma rotation is observed when an n=3 RMP is applied (see the solid line in Figure 7b). The MARS-Q code also predicts little braking in this plasma configuration (see the solid circles in Figure 7b). The braking torque due to the $\vec{j} \times \vec{B}$ force in the LSND discharge, with similar n=3 RMP, is calculated to be T=-0.36 Nm compared to -0.02 Nm for the CDN discharge. A possible explanation for this difference may lie in the fact that the CDN plasmas have a much larger $q_{95}$ ($q_{95}$ ~ 7.0 compared to 3.0 in the LSND), which means that there are more rational surfaces near the edge of the plasma which can screen the perturbation and hence reduce the $\vec{j} \times \vec{B}$ torque on the core of the plasma. Another difference between the two discharges is the normalised plasma pressure ($\beta_N$). The LSDN discharge has $\beta_N = 3.6$ while the CDN discharge has $\beta_N=3.1$, both of which are much lower than the ideal no-wall limit for n=3 modes for these discharges ($\beta_N^{no-wall}$~4.5). However, if the pressure is artificially increased in the CDN discharge to $\beta_N=3.8$ a kink-like response is predicted to be excited in the plasma and the MARS-Q code then predicts damping of the core rotation (see the open circles in Figure 7b). These possible explanations will be explored further in future modelling and experimental studies.

Previous stability analyses on MAST using the ELITE code [20] have suggested that reducing the sheared toroidal rotation at the edge of the plasma has a destabilizing effect on the peeling–ballooning modes [21]. This analysis showed that the rotation has the strongest stabilizing effect on the high-n modes and hence a reduction in rotation may lead to ELMs with a higher mode number. However, the experimental observation is that even in the case of the large core braking observed in the LSND experiments, very little change is observed in the plasma edge rotation [19]. Finally, no change in toroidal rotation velocity is observed during the mitigated phases of CDN discharges. This would suggest that changes in rotation are not the dominant explanation for the changes in ELM frequency.

Since the screening in MARS-F is mainly due to the toroidal rotation velocity, the braking predicted by the code and observed in the experiment for the LSND discharges, means that the penetrated field varies as a function of time. Figure 8 shows the radial profiles of the normalised resonant field component calculated in the vacuum approximation from ERGOS and including the plasma response and rotation screening using MARS-F. The MARS-F calculations have been performed for the LSND shot with the RMPs coils in the n=4 configuration using the experimentally measured initial toroidal (i.e. before the RMPs have been applied) and final saturated (i.e. after the rotation braking) rotation profiles. The n=4 configuration is chosen since this has the largest braking that attains a saturated level. For $\sqrt{\Psi_{pol}} < 0.9$ the value of $b^r_{res}$ increases by more than an order of magnitude but still remains two orders of magnitude lower than the vacuum calculations. However, the Dirichlet boundary condition used for these simulations with the MARS-Q code means that, similar to what is observed experimentally, the edge rotation velocity does not change and hence the value for penetrated field at the edge changes little.

### 4. Effect of RMPs on ELM size and pedestal characteristics

Figure 9a shows a plot of the energy loss per ELM ($\Delta W_{ELM}$), derived from the change in plasma stored energy, versus $f_{ELM}$ for natural and mitigated ELMs for the CDN and LSND discharges. In both cases the application of the RMPs produces an increase in $f_{ELM}$ and corresponding decrease in $\Delta W_{ELM}$ consistent with $f_{ELM}.\Delta W_{ELM}= 500$ kW (represented by the solid curve in Figure 9a). It is interesting to note that the two magnetic configurations have such a similar trend, presumably due to the fact that the two have similar plasma current and input power (sum of Ohmic and neutral beam), although the natural ELM frequency is lower and consequently the ELM energy loss is larger in the LSND discharges. The fact that $f_{ELM}.\Delta W_{ELM} \sim$ constant suggests that there is little change in the inter-ELM transport between the natural and mitigated phases. Not only does the total energy lost per ELM decrease but so does the number of particles. For both the LSND and CDN discharges the change in plasma density due to an ELM expressed as fraction of the pre-ELM pedestal density has a mean value of $\Delta n_e/n_e^{ped} = 0.04$ for the natural ELMs. The associated so-called "convected" ELM energy loss typically remains constant as the density and collisionality are varied in any given machine and configuration [1][22]. Therefore it is interesting to see that for the mitigated ELMs the fractional density loss decreases to 0.02 at the highest ELM frequencies (Figure 9b). In fact since the pedestal density has also decreased in the mitigated stages this means that the actual particle loss has been decreased more than a factor of 2. Since the pedestal temperature remains unaltered for the mitigated ELMs, models for the ELM energy loss based on parallel loss processes (see for example [23] and [24]) would require that the time over which the ELM energy loss occurred was reduced. It would be interesting to investigate this in future experiments.

In order to avoid damage to in-vessel components in future devices, such as ITER, it is the peak heat flux density at the divertor that is important rather than the actual ELM size. The heat fluxes at the LFS lower divertor have been measured using infrared thermography in both the CDN and LSDN discharges for the natural and mitigated ELMs. Figure 10a and b shows the peak heat flux density at the target ($q_{peak}$) as a function of $\Delta W_{ELM}$ for the LSND and CDN discharges respectively. In both the LSND and CDN discharge types the increase in ELM frequency and decrease in $\Delta W_{ELM}$ does lead to reduced heat fluxes at the target, although it also results in a smaller wetted area at the target meaning that the reduction in $q_{peak}$ is not as large as the reduction in $\Delta W_{ELM}$ [25]. In the case of the LSND discharges the mitigated and natural ELMs follow the same trend and show that a reduction of a factor of 3 in $\Delta W_{ELM}$ (i.e. from 15 to 5kJ) produces a reduction in $q_{peak}$ of 1.8 (from 15.8 to 9.4 MWm$^{-2}$). However, if this trend is linearly extrapolated to very small energies (i.e. $\Delta W_{ELM}\sim 0$) the predicted target heat flux density is 6.3 MWm$^{-2}$, which is much larger than the typical inter-ELM values of 1MWm$^{-2}$. Turning next to the CDN data, a similar reduction of a factor of 3 in $\Delta W_{ELM}$ (i.e. from 6 to 2kJ) produces a slightly larger reduction in $q_{peak}$ of 2.1 (from 9 to 4.3 MWm$^{-2}$). Although these data are at smaller energies than the LSDN data, a linear extrapolation to $\Delta W_{ELM}= 0$ would still indicate a peak heat flux of $\sim 2$ MWm$^{-2}$ for $\Delta W_{ELM}= 0$ that is larger that the typical inter-ELM peak heat fluxes of $\sim 0.5$ MWm$^{-2}$.

A study of the visible images observed during the ELMs in the LSND discharges shows that the filament structures have similar characteristics for both natural and mitigated ELMs [19]. Based on the toroidal mode number of the filaments it would appear that the mitigated ELMs still have all the characteristics of type I ELMs even though their frequency is higher and their energy loss is reduced.

The pedestal electron density and temperature characteristics have been measured using a Nd YAG Thomson Scattering (TS) system. The radial pedestal profiles have been fitted using a modified tanh function to determine the pedestal height, barrier position and width on both the high field side (HFS) and low field side (LFS) of the plasma as a function of time in the ELM cycle. Previous analyses of the inter-ELM pedestal evolution on MAST have shown that the pressure gradient is effectively constant during the inter-ELM period with the pressure pedestal width growing as the pedestal height increases [26] . As the pedestal width increases the peeling-ballooning pressure gradient boundary decreases leading to the triggering of the ELM [26]. Figure 11a shows the evolution of the pressure pedestal height ($P_e^{ped}$) as a function of time after an ELM for a LSND discharge without and with RMPs in an n=6 configuration. The evolution is similar in the shots with and without RMPs; however, in the shots with RMPs applied the ELM is triggered earlier in the cycle at a lower value of $P_e^{ped}$, reflecting the increased ELM frequency. In the shot with RMPs applied the pressure pedestal width ($\Delta P_e$) is also larger throughout the ELM cycle (see Figure 11b), which results in an even larger decrease in the peak pressure gradient. For the CDN discharges the increase in the pedestal width is not so large when the RMPs are applied [27].

A stability analysis has been performed on these discharges using the ELITE stability code [20]. Figure 12a shows the stability boundary and the experimental point in a plot of peak edge current density ($j_\phi$) versus normalised pressure gradient ($\alpha$) for a LSND discharge without RMPs and with the RMPs in an n=6 configuration (the n=4 configuration gives a similar result). The results show that for the discharge without RMPs the experimental point lies in the region unstable to peeling-ballooning modes, a trait often associated with type I ELMs. For the shot with RMPs the broader pedestal width does decrease the critical pressure gradient required for peeling-ballooning instabilities. However, the reduction in the experimental pressure gradient when the RMPs are applied is larger than this reduction in stability boundary meaning that the experimental point is significantly below the new critical threshold.

Figure 12b shows a similar plot for the CDN discharge without and with RMPs in a n=3 configuration. In this case because the change in pedestal width is small the stability boundary changes little and when the RMPs are applied the experimental point clearly sits in a region stable to peeling-ballooning modes. Therefore it is not apparent why the ELM frequency should be higher in the discharges with RMPs applied.

Such arguments assume toroidally and poloidally symmetric measurements and that smooth edge flux surfaces are maintained. However, measurements with the TS system show that while the application of the RMPs produce a similar change in the height of the density pedestal on the HFS and LFS of the plasma a noticeable difference is observed in both the barrier location and pedestal width [27]. On the HFS the barrier location remains unchanged while on the LFS of the plasma, the barrier position is displaced radially inwards or outwards depending on the type of RMP applied and phase relative to the viewing diagnostics. The displacement, which is measured at z=0, is largest in the case of the application of the RMPs in an n=6 configuration to the LSND discharges where the transport barrier location is shifted by 20 mm [27]. In the CDN discharges with the coils in an n=3 configuration, the displacement varies between 6 and 10 mm depending on the phase of the applied perturbation [27].

As an example of how these displacements could be produced calculations have been performed using the ERGOS code by tracing field lines originating from the mid-plane region of the plasma through the 3D vacuum field. Each field line is traced for 200 toroidal turns or until it reaches the divertor target. The calculations include fields from the RMP coils, a parameterisation of the intrinsic error fields and the fields due to the ex-vessel error field correction coils  The minimum normalized flux that the field line experiences during its trajectory is recorded. Figure 13a and b show the vacuum magnetic field structure when an n = 6 RMP is applied to a LSND discharge and an n=3 RMP is applied to a CDN discharge respectively. The distance is plotted relative to the original separatrix position. As observed experimentally, the displacement in the LSND discharge is larger than in the CDN discharge; however, it is not possible to quantify the edge displacement accurately. If a given value of temperature and density could be associated values of $\Psi_N^{min}$, then Figure 13 would suggest that the pedestal width and gradient would vary as a function of toroidal angle. There is preliminary evidence that such a variation is observed experimentally in the CDN discharges where the measured pedestal width and gradient is different for the two phases of the applied perturbation [27]. Since the difference is predicted to be largest for the application of the RMPs in an n=6 configuration to the LSND discharges, this will be investigated in future experiments by applying the perturbation in two phases. As was discussed in [28] the application of 3D fields to the plasma also produces changes in the underlying equilibrium and these 3D equilibrium effects will also need to be taken into account in future modelling.

### 5. Observation of Lobes at the X-point

The lower X-point region of the plasma has been viewed using a toroidally viewing camera with a spatial resolution of 1.8mm at the tangency plane. The image has been filtered with a $He^{1+}$ (468 nm) filter and the images obtained using an integration time of 2 ms. This line has been chosen since it is well localised in the separatrix region for the typical plasma conditions found in MAST. Figure 14a shows one such image obtained during an inter-ELM period for a shot with $I_{ELM}$=5.6 kAt with the coils in an n=6 configuration. Clear lobe structures are seen near to the X-point. In an ideal axi-symmetric poloidally diverted tokamak the magnetic separatrix (or LCFS) separates the region of confined and open field lines. The idea that so-called "Manifold" structures could exist was probably first introduced to the tokamak community by Evans et al., [29][30]. Non-axi-symmetric magnetic perturbations split this magnetic separatrix into a pair of so called "stable and unstable manifolds" [29][31]. Structures are formed where the manifolds intersect and these are particularly complex near to the X-point. The manifolds form lobes that are stretched radially both outwards and inwards. Some of these lobes can intersect the divertor target and result in the strike point splitting often observed during RMP experiments [32][33].

Calculations of what these lobes should look like have been performed based on numerical field line tracing using the ERGOS code (see Figure 14b). When the modelled data is mapped onto the image taking into account the tangency location, a good quantitative agreement is obtained between the number and separation of the lobes, however, there appears to be a discrepancy in their radial extent [34], with the experimental images being shorter. This could be due to: 1) the sensitivity to the distribution of the $He^{1+}$ emission, 2) plasma screening of the applied fields and 3) cross field diffusive transport.

To investigate the effects of the $He^{1+}$ emission a forward model of the camera data has been constructed. Data recorded from a shot without RMPs were used to generate synthetic images by assuming that the $He^{1+}$ light is a flux surface quantity, and finding a light distribution function that provided a good match between the experimental and synthetic camera data. The ERGOS code was used to follow field lines in the 3D region monitored by the camera to determine the average magnetic flux along these field lines, which was then used to determine the light emission within the lobes. The resulting simulated image, which is shown in Figure 15a, shows that the radial extent of the simulated lobes is again too large.

To investigate the effects of screening, an ideal plasma response has been assumed and the helical currents required to screen the field at the rational surfaces have been introduced into the ERGOS code, using the method described in reference [35]. The screening currents have been calculated at each rational surface from the core out towards the edge of the plasma. The simulations have been carried out where the furthest out screening current is located at the flux surface with poloidal mode number m=12, 15 or 18 corresponding to q = 2, 2.5 and 3 (for the n = 6 configuration of the RMPs used here). These surfaces are located at $\sqrt{\Psi_{pol}}$ = 0.929, 0.962 and 0.980 respectively. The aim of these simulations is to estimate how far the perturbations have penetrated into the plasma. The open symbols in Figure 16 show the radial profile of the normalised radial resonant field component ($b^r_{res}$) calculated in the vacuum approximation from ERGOS, for the cases where no screening has been used (circle) and cases where the furthest out screening current is located at m = 12 (diamond), 15 (square) and 18 (triangle). As can be seen the screening currents effectively zero the perturbation inside their radial location and affect the region outside. In the case of the screening currents out to m=18 they reduce the edge field by a factor of 5.

Simulated images of the lobes have been generated for each of the screening currents. The case where the furthest out current is at m=12, which would imply a penetration of the field to $\Psi_{pol}$=0.86, produces lobes that are too large. The simulations for m=15 and m=18 are more similar, with the m=18 simulations giving the best match to the observed image  The resultant simulated image using screening currents out to m=18, corresponding to $\Psi_{pol}$=0.96, is shown in Figure 15b. The radial extent and width of the lobes is in good agreement with the experimental image shown in Figure 14a. This suggests that the field does penetrate the plasma edge but only up to $\Psi_{pol}$=0.96, which corresponds to a location just inboard of the pedestal top. Calculations have also been performed using the MARS-F code for the same discharge. The calculated $b^r_{res}$ profile, taking into account the plasma response and screening due to the saturated level of the toroidal rotation (i.e. using the braked rotation profiles), is superimposed on Figure 16 as the solid squares. The results from MARS-F are in good agreement with the results of the ad hoc screening model when the last screening current is at m=18.

The radial extent of the lobes has been measured for repeat shots performed at different values of $I_{ELM}$ [19]. For coil currents above a threshold ($I_{THR}$) the extent of the lobes increases approximately linearly with $I_{ELM}$-$I_{THR}$. Hence a clear correlation is observed between the size of the lobe length and the change in ELM frequency, which may suggest that the lobes themselves are having a direct impact of the stability of the edge plasma to peeling ballooning modes.

### 6. Summary and discussion of implications for ELM mitigation

Sustained ELM mitigation has been achieved using RMPs with a toroidal mode number of n=4 and n=6 in lower SND and with n=3 in CDN plasmas on MAST. The ELM frequency increases by up to a factor of eight with a similar reduction in ELM energy loss. A threshold current for ELM mitigation is observed, which depends on the configuration of the applied RMP. Above this threshold the ELM frequency increases approximately linearly with current in the coils. Calculations have been performed in the vacuum approximation and taking into account the plasma response; in both cases above a threshold value the increase in ELM frequency scales linearly with the RMP amplitude ($b^r_{res}$), however, it is not possible to explain the differences in the threshold value. A comparison of the filament structures observed during the ELMs in the natural and mitigated stages show that the mitigated ELMs still have all the characteristics of type I ELMs even though their frequency is higher, their energy loss is reduced and the pedestal pressure gradient is decreased.

Although it has not been possible to identify a single parameter that determines the extent and onset of ELM mitigation several changes to the characteristics of the plasma have been observed that could explain the increases in ELM frequency. Firstly, in the LSND plasmas the application of the RMPs produces substantial braking of the core rotation and this could have an effect on the ELM stability. However, as was discussed in section 3, the rotation at the edge pedestal is little affected by the RMPs. In addition, as was reported in [19], LSND shots have been repeated at a reduced outer radius, which have a smaller edge value of $b^r_{res}$ and no effect on the ELM frequency, but they still had substantial braking of the core toroidal rotation. This combined with the fact that ELM mitigation is observed in the CDN shots, which do not have core braking, would suggest that changes in rotation are not the dominant reason for ELM mitigation.

The second effect observed are the changes in the pedestal characteristics. The midplane LFS of the plasma acquires a 3D perturbation. This appears to produce regions of increased and decreased pressure gradient. Similar to what is observed in the modelling of pellet induced ELMs [36], toroidally and poloidally localised increases in the pressure gradient may be responsible for triggering the ELMs at what, in a symmetric configuration, would be a stable point. In addition, as was discussed in [28] the effect the 3D fields have on the plasma equilibrium has been investigated using the VMEC code [37]. These calculations predict a peak to peak displacement of ~5 cm, in good agreement with that measured experimentally. The influence of the 3D corrugation on infinite-n ballooning stability has been examined using the COBRA code [38]. The growth rate of the n=$\infty$ ballooning modes at the most unstable toroidal location is a factor of two larger than the axisymmetric case i.e. the plasma edge is strongly destabilised at certain toroidal positions.

Finally, lobe structures have been observed in the X-point region and the appearance and size of the lobes is correlated with the increase in ELM frequency. To examine the effect these lobes may have on the edge stability, in reference [39] an axisymmetric stability analysis has been performed using the ELITE code [20]. A degradation in the ballooning stability was observed as the lobe size was increased, which originated from the perturbed field lines dwelling in the region of unfavourable curvature due to the presence of the lobes. In order to do this correctly the 3D lobe structures need to be taken into account in a 3D code. The calculations performed in [39] highlight the difference between LFS lobes, which lead to a destabilisation of the ballooning stability compared to HFS lobes which cause a stabilisation. This may help to explain why a higher increase in ELM frequency has been achieved in CDN discharges compared to LSND discharges, since in a CDN discharge the lobe structures only exist on the LFS of the plasma (see Figure 17 ) whereas in a LSND discharge the lobes exist on both the HFS and LFS of the plasma (Figure 14b).

The observation of lobe structures and the associated modelling allow an estimate to be made of the field penetration. The field screening is found to be in good agreement with the calculations from MARS-F, which is a single fluid model. It does not capture the subtle physics of 2-fluid effects, which mainly relate to diamagnetic flow effects. In order to fully model the physics of the pedestal region a full nonlinear two-fluid model is most likely required [40][41][42]. In particular, it has been suggested [43][44] that the RMPs are screened due to the perpendicular rotation of the electron fluid and that for a low resistivity (or ideal) plasma the RMP field will only be large close to rational surfaces where the total electron perpendicular velocity ($V_e^{\perp}$) is near zero. To test if this criterion is approximately met in the MAST plasmas the $\vec{E} \times \vec{B}$ velocity and the electron diamagnetic velocity have been estimated for the LSND discharges for which the field penetration has been estimated. Although there are uncertainties in these quantities, as can be seen from Figure 18, the total electron rotation velocity is consistent with having a zero inside the plasma near to the pedestal region ($V_e^{\perp}$=$V_{ExB}$+$V_e^*$ crosses zero between $\psi_{pol}$ = 0.935 and 0.955), which considering the uncertainties is in good agreement with the maximum RMP penetration depth estimated form the lobe analysis ($\psi_{pol}$ = 0.96). However, in the region between this point and the edge of the plasma (i.e. 0.96 <$\psi_{pol}$ < 1.0) $V_e^{\perp}$ is large and it is interesting to understand how the fields can penetrate through this region. As was discussed in references [45] and [46] one possible reason why the RMP field may be less screened in this region is because the resistivity is large and hence the screening currents are reduced. These effects are observed in the two fluid resistive MHD modelling [42].

The results presented in this paper suggest that ELM mitigation due to RMPs results from the 3D perturbations to the separatrix, which then cause a degradation of the edge stability to peeling ballooning modes. As depicted in Figure 19, in a natural ELM cycle the pressure pedestal height and width increase until the peeling ballooning limit is reached. The application of the RMPs leads to a 3D corrugation of the mid-plane separatrix leading to a pressure gradient that is no longer axisymmetric, which combined with the lobe structures near to the X-point, leads to a decrease in the stability boundary [39]. The inter-ELM transport appears to be the same in the natural and the mitigated ELMs meaning that the pressure profile reaches this new, lower stability limit earlier in the natural ELM cycle and hence an increase in ELM frequency results. The level of the ELM mitigation achieved would then depend on the location of this new stability limit. The price paid for the mitigated ELMs, however, is a reduction in the maximum pedestal height achieved. In order to achieve ELM suppression a mechanism would then need to be found to stop the pedestal evolving towards the stability boundary. In fact recent findings on DIII-D suggest that the RMPs may induce an island at the top of the pedestal and the transport due this island impedes the widening of the pedestal, which then stops the peeling ballooning limit being reached [47]. The optimum would be to arrange things in such a way that this was achieved with the minimum reduction in pedestal height and hence plasma performance.

## Acknowledgement

This work was funded partly by the RCUK Energy Programme under grant EP/I501045 and the European Communities under the contract of Association between EURATOM and CCFE. The views and opinions expressed herein do not necessarily reflect those of the European Commission. This work was also part funded by the grant agency of the Czech Republic under grant P205/11/2341. This work was carried out within the framework of the European Fusion Development Agreement.

**Figures**

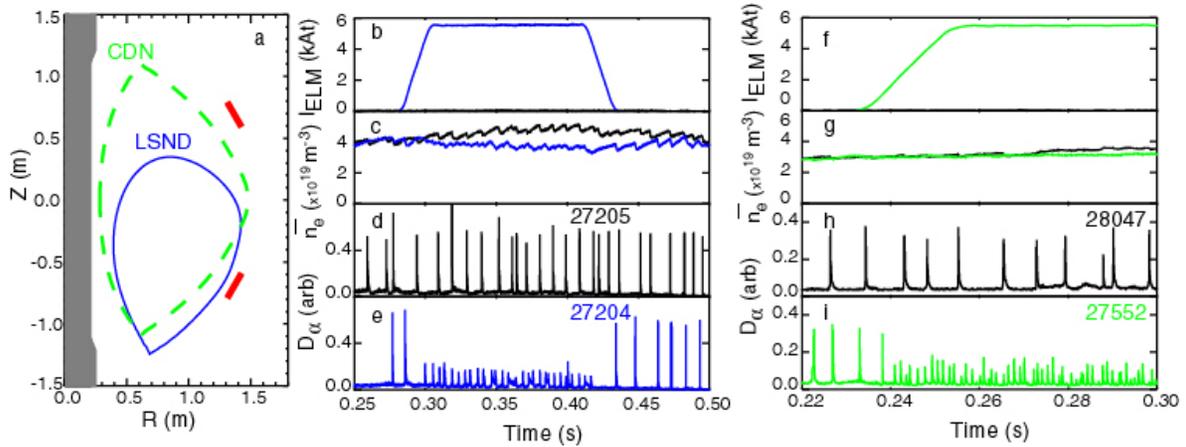

**Figure 1** a) Poloidal cross section of the LSND (solid curve) and CDN (dashed) plasmas used together with the location of the centre column and ELM coils. For the LSND discharge time traces of b) the current in the ELM coils ($I_{ELM}$) c) line average density and the target $D_\alpha$ intensity for discharges d) without and e) with an n=6 RMP from the lower row of coils. For the CDN discharge time traces of f) the current in the ELM coils ($I_{ELM}$) g) line average density and the target $D_\alpha$ intensity for discharges h) without and i) with an n=3 RMP from both rows of coils in an even parity configuration.

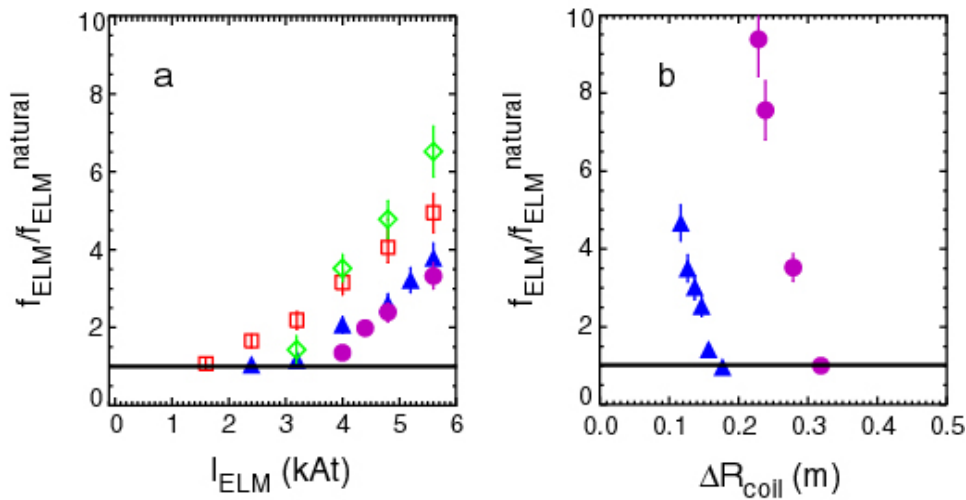

**Figure 2** ELM frequency divided by the natural ELM frequency a) versus ELM coil current ($I_{ELM}$) at fixed distance between the plasma and the coils ($\Delta R_{coil}$) and b) versus $\Delta R_{coil}$ at IELM = 5.6 kAt with the RMPs for SND discharges in an n= 4 (squares) and n=6 (triangles) and the CDN discharges in an n= 3 even (circles) and 90L (diamond) configuration.

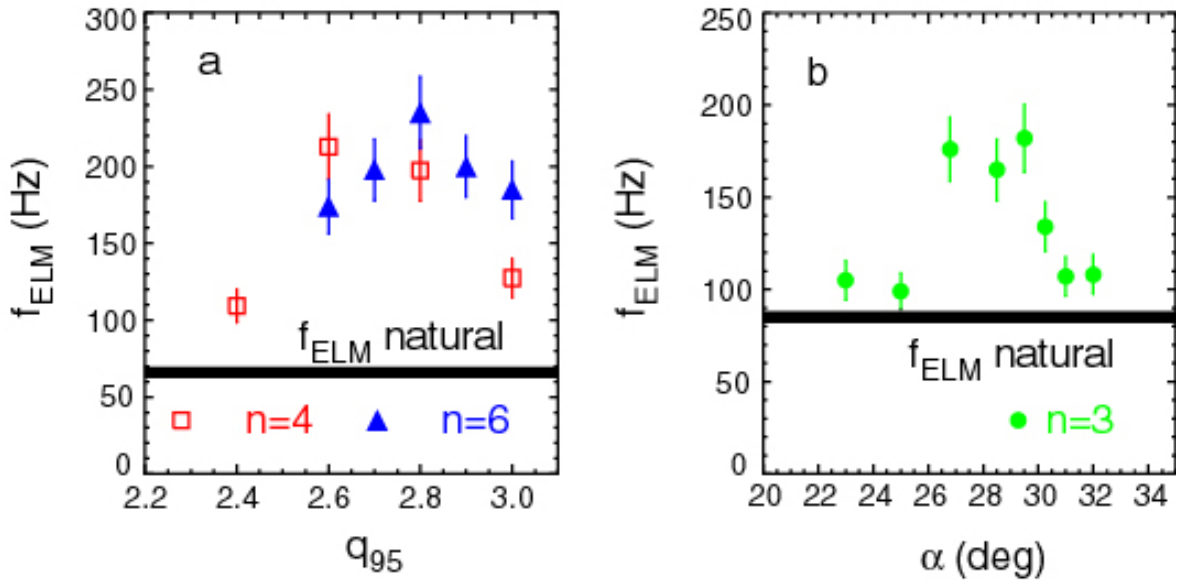

**Figure 3** ELM frequency ($f_{ELM}$) as a function of a) $q_{95}$ for LSND shots with $I_{ELM}$ =4.0 kAt in the RMPs in an n=4 (open squares) and $I_{ELM}$ =5.6 kAt in an n=6 (closed triangles) configuration from the lower row of coils and b) the pitch angle ($\alpha$) of the applied RMP field in an n=3 configuration from both rows of coils applied to CDN discharges.

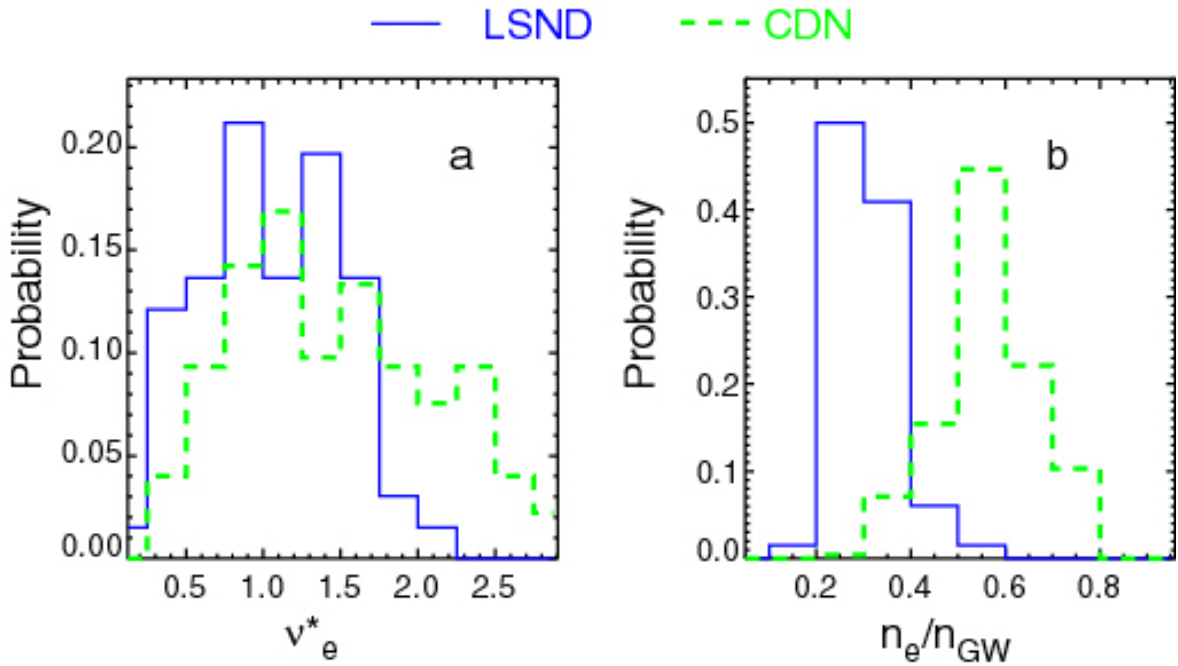

**Figure 4** Probability distribution of a) the pedestal collisionality ($\nu^*_e$) and b) the line averaged density as a fraction of the Greenwald density ($n_e/n_{GW}$) for the LSND (solid) and CDN (dashed) discharges.

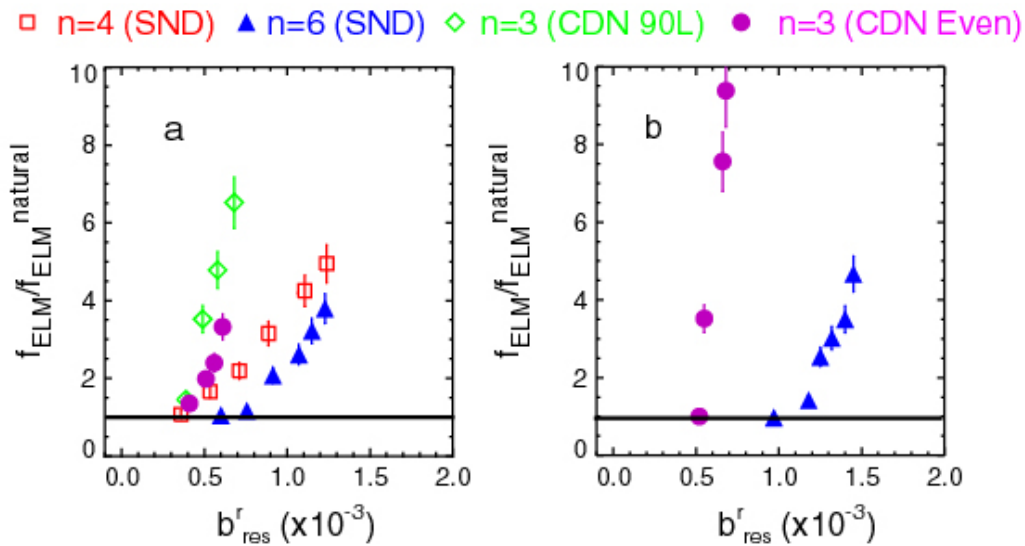

**Figure 5** ELM frequency ($f_{ELM}$) as a function of maximum resonant component of the applied field ($b^r_{res}$) resulting from a) a scan in $I_{ELM}$ and b) a scan in $\Delta R_{coil}$.

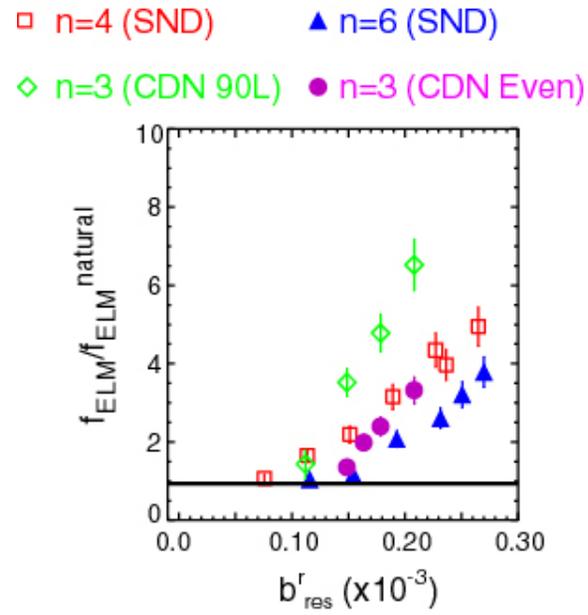

**Figure 6** Results from a scan of $I_{ELM}$: normalised $f_{ELM}$ as a function of the maximum resonant component of the applied field ($b^r_{res}$) calculated taking into account the plasma response.

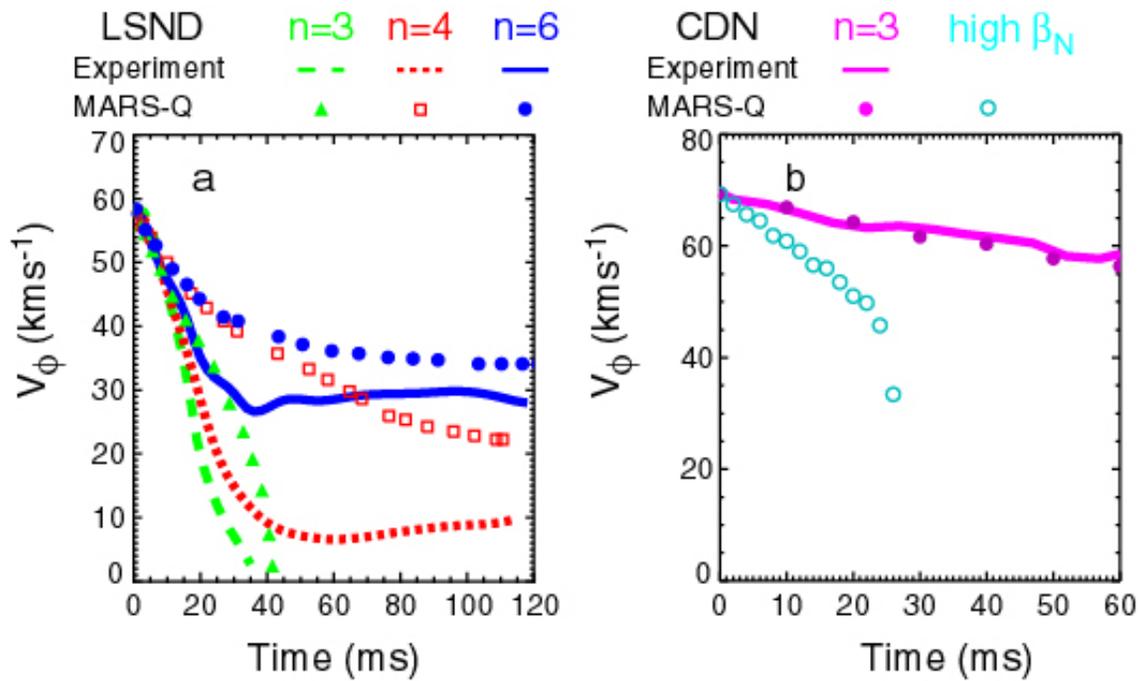

**Figure 7** The experimentally measured core toroidal rotation velocity (lines) as a function of time after which the RMPs reached flat top (Δt)and the results from the MARS-Q code simulations (symbols) for a) LSND discharges and b) CDN discharges.

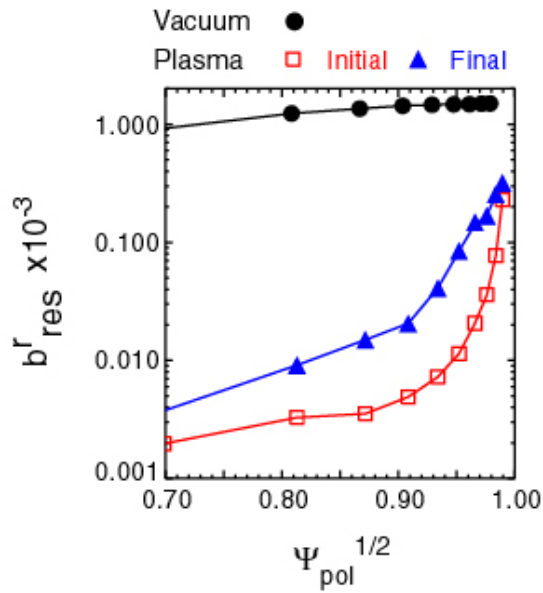

**Figure 8** Radial profile of the normalised resonant component of the applied field ($b^r_{res}$) for a LSND discharge with the RMPs in an n=4 configuration calculated in the vacuum approximation (circle) and taking into account the plasma response using the toroidal rotation profiles before the RMPs are applied (squares) and after the braking (triangles).

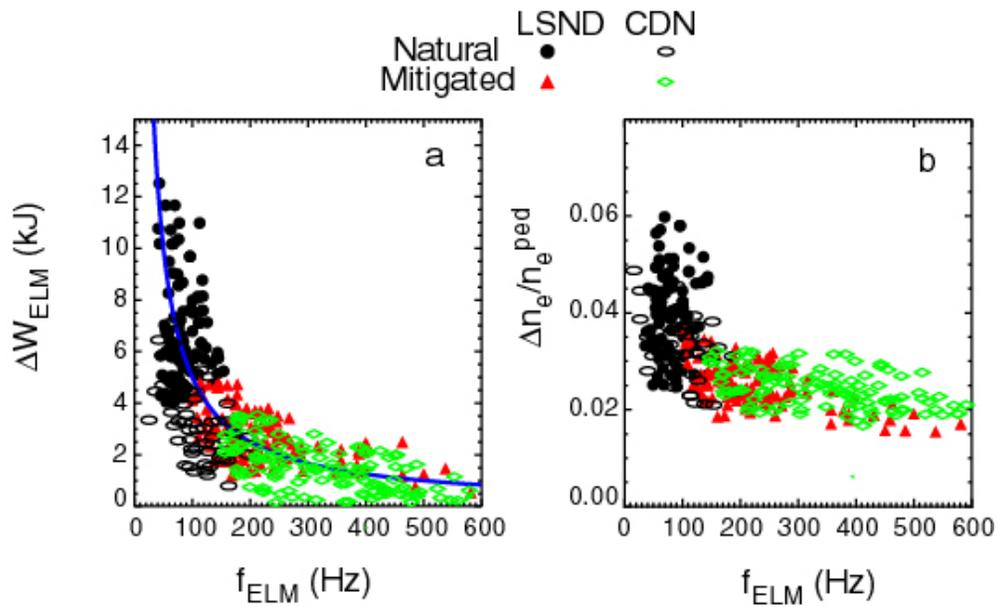

**Figure 9** a) ELM energy loss ($\Delta W_{ELM}$) and b) ELM particle loss expressed as a fraction of the pedestal density ($\Delta n_e/n_e^{ped}$) as a function of ELM frequency ($f_{ELM}$) for natural ($I_{ELM}=0$ kAt) in LSND (solid circle) and CDN (open oval) and mitigated ELMs in LSND (triangle) and CDN (diamond).

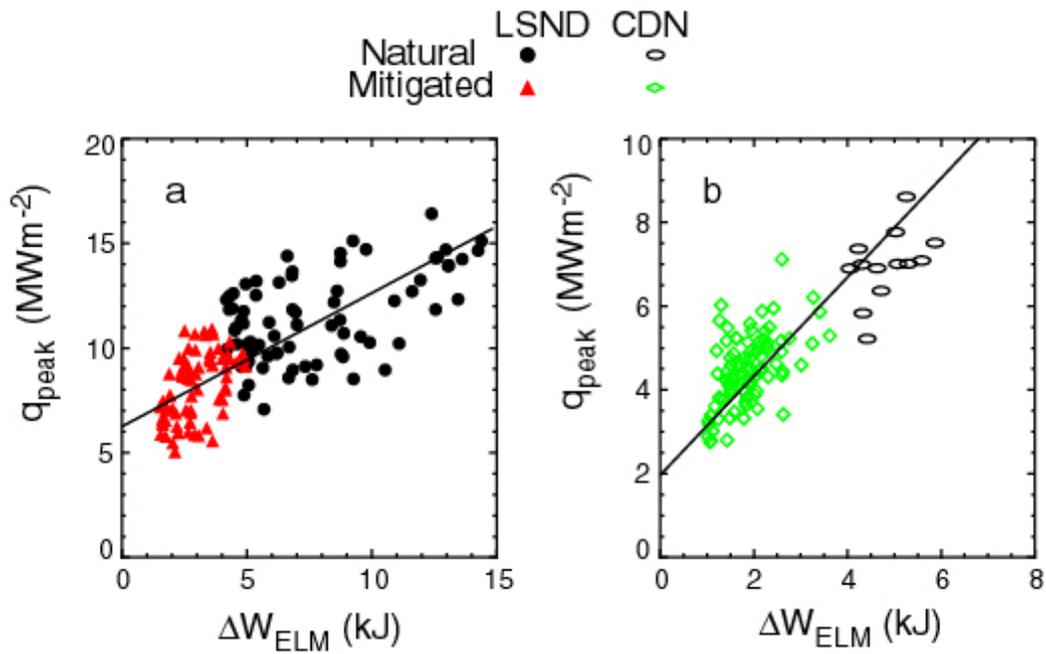

**Figure 10** Maximum peak heat flux ($q_{peak}$) during an ELM at the low field side divertor as a function of $\Delta W_{ELM}$ in a) a LSND discharge for natural ($I_{ELM}$=0 kAt) (solid circle) and mitigated (triangle) triangle and b) a CDN discharge for natural (open oval) and mitigated (diamond) ELMs. Superimposed on the plots are lines showing linear fits to the data.

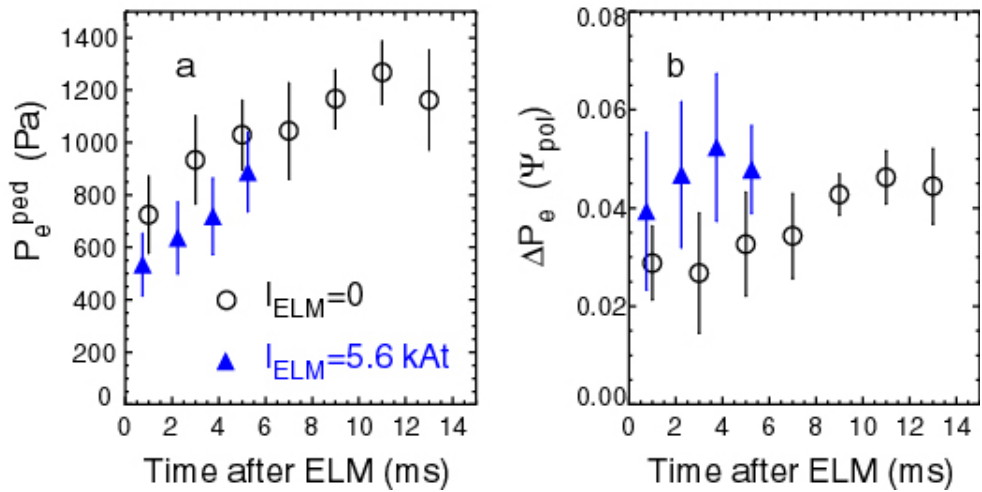

**Figure 11** Evolution of the electron pressure pedestal  a) height and b) width during the ELM cycle for shots without (circles) and with (triangles) RMPs in an n=6 configuration.

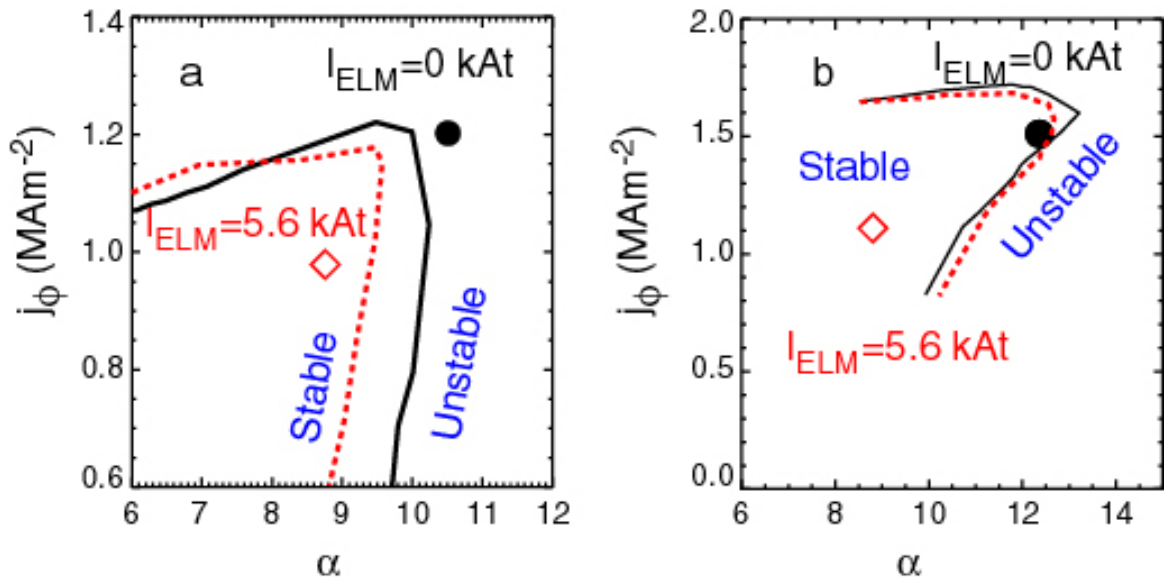

Figure 12 Peeling-ballooning stability diagram in the edge current density ($j_\phi$) versus normalised pressure gradient ($\alpha$) plane, calculated for shots with $I_{ELM} = 0$ kAt (solid line) and $I_{ELM} = 5.6$kAt in the RMPs (dashed line) in a) an LSND configuration with n=6 and b) a CDN configuration with n=3.  The circle and diamond represent the experimental points for the $I_{ELM} = 0$ and 5.6 kAt cases respectively.

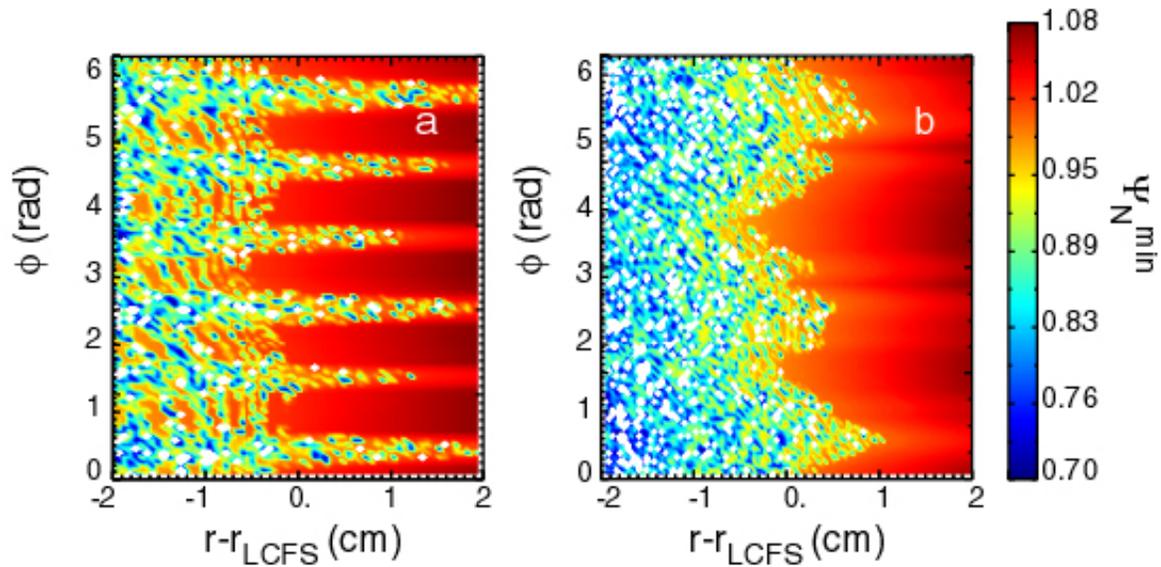

**Figure 13** A two-dimensional contour plot of the minimum flux surface that each field lines starting at the LFS midplane (z=0) experiences during its trajectory as a function of distance from unperturbed last closed flux surface (r-r$_{LCFS}$) and toroidal angle (ϕ) for a) a LSND discharge with the RMPs in an n=6 configurations and b) CDN for the coils in a n=3 even parity configuration.

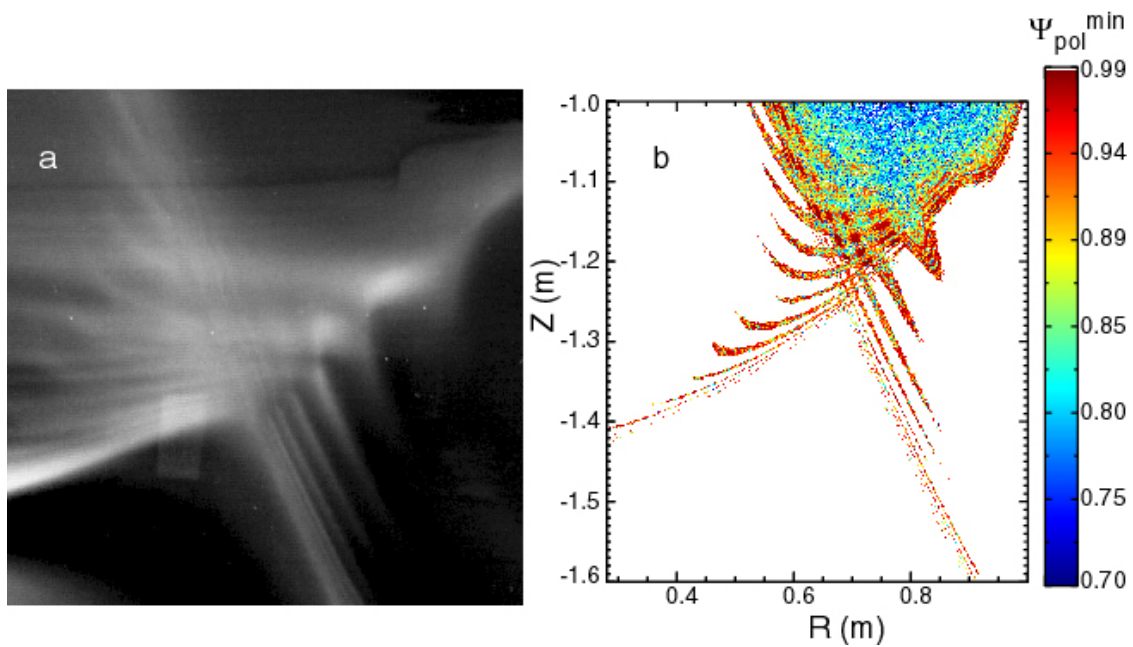

**Figure 14** A LSND discharge with RMPs in an n=6 configuration a) image of the He$^{1+}$ emission from the X-point region captured during an Inter-ELM period of a LSND H-mode with the RMPs and b) Poincare plots from ERGOS showing the minimum value of $\Psi_{pol}$ reached by a field line.

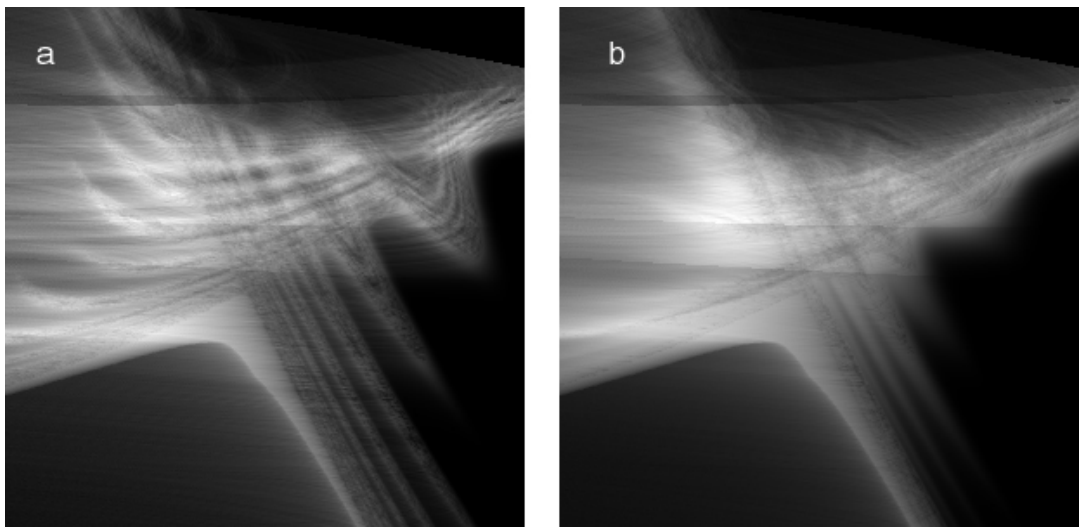

**Figure 15** Simulation of the He1+ light emission for a LSND discharge using a) unscreened and b) screened RMPs in a n=6 configuration.

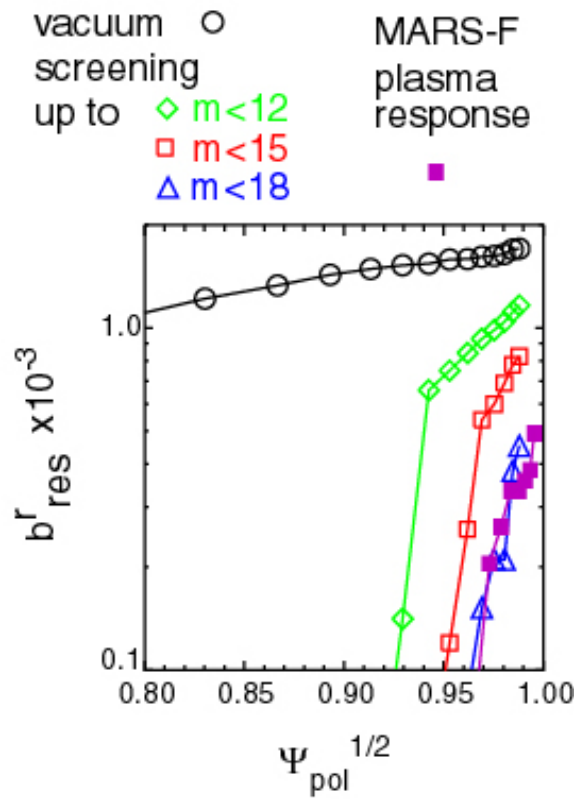

**Figure 16** Calculated profiles of the normalised resonant component of the applied field ($b^r_{res}$) produced with 5.6 kAt in the ELM coils in an n=6 configuration applied to a LSND discharge using the vacuum approximation with no screening (open circle) and including screening from an ad hoc model out to m=12 (diamond), m=15 (open squares) and m=18 (triangle). Also shown is the profile taking into account the plasma response using the MARS-F code (closed squares).

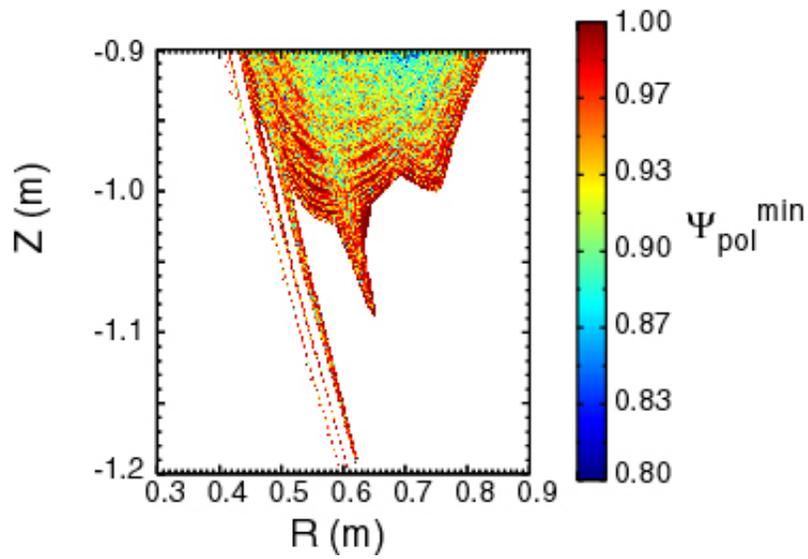

**Figure 17** Poincare plots from ERGOS showing the minimum value of $\Psi_N$ reached by a field line for a CDN discharges with RMPS in an n=3 even parity configuration.

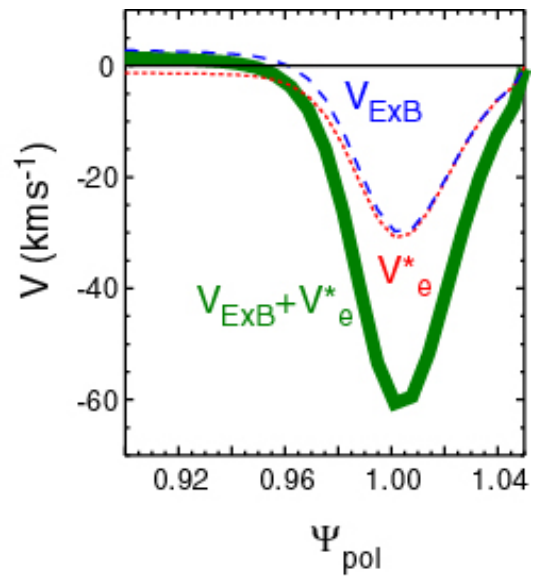

**Figure 18** Profiles of the $\vec{E} \times \vec{B}$ rotation of plasma ($V_{ExB}$) (dashed), the electron diamagnetic rotation $V_e^*$ (dotted) and the total perpendicular rotation velocity of the electrons ($V_e^{\perp} = V_{ExB} + V_e^*$) (solid).

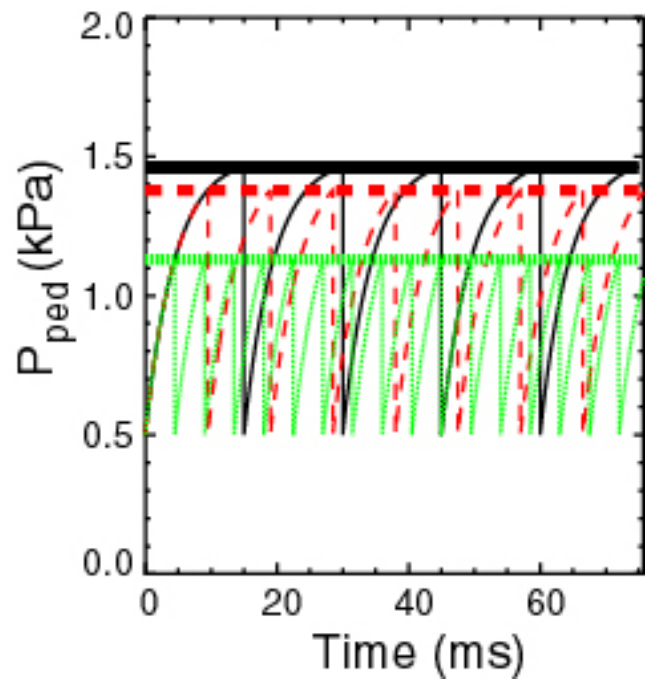

**Figure 19** Cartoon depicting the evolution of the pressure pedestal (curves) and the ballooning stability limit (horizontal line) during the ELM cycle for natural (solid) and various levels of mitigation (dotted and dashed).